# Object Sensing for Fruit Ripeness Detection Using WiFi Signals

Sheng Tan, Linghan Zhang, Jie Yang
Florida State University, Tallahassee, Florida, USA

*Abstract*—This paper presents FruitSense, a novel fruit ripeness sensing system that leverages wireless signals to enable non-destructive and low-cost detection of fruit ripeness. Such a system can reuse existing WiFi devices in homes without the need for additional sensors. It uses WiFi signals to sense the physiological changes associated with fruit ripening for detecting the ripeness of fruit. FruitSense leverages the larger bandwidth at 5GHz (i.e., over 600MHz) to extract the multipath-independent signal components to characterize the physiological compounds of the fruit. It then measures the similarity between the extracted features and the ones in ripeness profiles for identifying the ripeness level. We evaluate FruitSense in different multipath environments with two types of fruits (i.e, kiwi and avocado) under four levels of ripeness. Experimental results show that FruitSense can detect the ripeness levels of fruits with an accuracy over 90%.

## I. INTRODUCTION

Recent advances in wireless technology have greatly expanded the WiFi usage from providing laptop connectivity to connecting mobile and smart devices to the Internet and home networks. Such an evolution has resulted in the prevalence of WiFi devices, which provides opportunities to extend WiFi's capabilities beyond communication, particularly in human sensing. As the wireless signals travel through space, they interact with human body and undergo wave phenomena such as reflection and diffraction. These phenomena lead to multipath effects, which carry a rich set of information about the physical environment including the human location and activities. Indeed, there has been growing interest within the wireless and mobile computing communities in using multipath effects to perform human sensing, ranging from large scale movements (e.g., daily activities [48], [50], [14], body movements [5], [37], [49], [54], hand gestures [26], [37]), to small scale motions (e.g., finger gestures [9], [43], [42], lip motions [47], breathing and heartbeats [30], [8], [14], [29] and location [55], [18], [51], [46], [57], [56], [28].

In this work, we further expand the WiFi sensing capabilities from human sensing to sensing bio-information of fruit crops. In particular, we seek to sense the degree of ripeness in fruits with WiFi signals. Monitoring the ripeness of fruit provides many benefits for several end-users, ranging from farmers, to distributors, to retailers and consumers. It can help farmers to determine the optimal time to harvest fruit as the quality heavily depends on when they are harvested. It can also assist fruit distributors in performing rapid sorting in storage facilities and deciding when to send their stock during post-harvest period. Moreover, retailers can minimize losses and maintain fruit quality through effective selling strategies based on accurate categorization of the fruit ripeness. Monitoring the ripeness of fruits throughout the supply chain thus can reduce waste as well as improve the consistency of quality for consumers. Consumers also can benefit, for example, knowing the degree of ripeness can avoid the unpleasantness of tart or rotten fruit.

Existing approaches for inferring the ripeness of fruit mainly depend on penetrometer [16], refractometer [41], or spectrometer [58]. As the firmness and sugar content are two most widely used indicators of fruit ripeness, the food industry primarily relies on penetrometer and refractormeter in the field for the purpose of ripeness detection [31]. While the penetrometer measures fruit firmness by quantifying the force required to insert a probe into the fruit [16], refractometer (e.g., Brix [17]) analyzes sugar content of the juice using light refraction. Both methods, however, are destructive and less acceptable to consumers. In contrast, spectrometer based approach is non-destructive. It splits light signals into a fruit and then measures the light that is emitted, absorbed or scattered by the fruit for ripeness inference [58]. However, traditional spectrometer instruments are bulky and expensive (i.e., costing several thousands of dollars) and are limited to controlled laboratory settings [38].

More recent work includes using advanced imaging techniques and ultrasonic measurement systems to analyze the image features (e.g., color and texture) and measure the ultrasonic attenuation of the fruit for ripeness detection [32], respectively. However, these methods require specialized equipments, and are less accessible to ordinary consumers. Recently, due to advances in materials and fabrication techniques, portable spectrometers that work together with smartphones have been realized. For example, Das *et al.* [13] propose a smartphone based spectrometer that can measure UV fluorescence of Chlorophyll found in apples, whereas the company Consumer Physics develops the sensor SCiO that can be integrated with smartphones to analyze the molecular composition of food [2], [1]. Although these portable solutions can be adapted by consumers, there also are non-negligible costs incurred in purchasing dedicated spectral sensors (i.e., at around $200).

In this paper, we introduce FruitSense, a new method for inferring the ripeness of fruit that is both non-destructive and low-cost. FruitSense uses WiFi signals to sense the physiological changes associated with fruit ripening for ripeness detection. It enables users to reuse off-the-shelf WiFi devices for ripeness monitoring, and thus can benefit a large number of users, such as farmers, retailers and customers.

In particular, fruit ripening involves a series of physiological changes leading to the development of a soft edible ripe fruit. Take avocado as an example, there are changes in total dry matter and moisture content during maturation and ripening. As moisture content decreases, the dry matter increases accordingly. When wireless signals travel through a fruit, the changes in the physiological compounds of the fruit during ripening lead to distinct and measurable effects on the received signals. We thus are able to infer the ripeness level based on the physiological changes interpreted by the received signals.

Accurately discerning the ripeness of fruit is challenging, however, when using a single pair of off-the-shelf WiFi devices. First, the measurable changes in the received signals due to fruit ripening are subtle, and the impact of fruit size on the received signals might distort such changes. To address these issues, we leverage frequency diversity by probing the fruit with WiFi signals at multiple frequency channels. Leveraging frequency diversity not only provides a rich set of information to capture the subtle changes, but also enables us to examine the change pattern among multiple channels, instead of a single channel, to mitigate the impact of fruit size.

Second, the signal propagation is dominated by multipath in typical indoor environments when the line-of-sight (LOS) is blocked by the fruit that is located between the transmitter and the receiver during the ripeness sensing process. While the WiFi based human sensing primarily relies on the signal reflections, the reflected signals from surroundings represent interferences to the signal component indicating the ripeness of fruit. This is because once the indoor environment changes, the changes in the received signals mainly reflect the differences in multipath propagation instead of the physiological change of the fruit. To accurately capture the changes corresponding to fruit ripening, we propose to isolate the signal component traveling through the fruit directly, from the ones reflected with longer paths. Specifically, we leverage the larger bandwidth at 5GHz (i.e., over 600MHz) to derive fine-grained power delay profile for multipath removal.

Third, due to unavoidable residual synchronization errors of the off-the-shelf WiFi NICs, the sampled phase of the channel frequency response contains linear phase errors, which result in significant distortion in the derived power delay profile. To calibrate the phase errors, we utilize the fact that the power delay profile derived from each individual channel should be identical as long as the probing packets across these channels were sent out within coherence time. We thus propose to formulate the phase error calibration as an optimization problem, in which we search for the best phase compensation that minimizes the differences in derived power delay profiles across multiple channels.

More specifically, our system first probes the fruit with WiFi signals hopping at all available 5GHz channels. The sampled channel frequency response, which is exported by the WiFi NICs in the form of Channel State Information (CSI), then goes through the calibration process to correct errors due to hardware limitation of WiFi NICs. We then stitch together the calibrated CSI measurements from each individual channel. As the usable WiFi channels in 5GHz are unequally and non-contiguous spaced, the inverse non-uniform Discrete Fourier Transform (NDFT) is used to derive the fine-grained power delay profile for multipath removal. Next, we identify and extract the signals that directly went through the fruit for ripeness detection. At last, we use the Maximal Overlap Discrete Wavelet Transform (MODWT) to extract signal features over multiple channels and compare the features against known ripeness profiles that identify the degree of ripeness.

We evaluate FruitSense in different multipath environments with two types of fruits: *kiwi* and *avocado*. The fruit samples are purchased from two different venders with different levels of ripeness. The volume of each type of fruit is about 75 for each level of ripeness. We also experience the scenarios where people are moving around and serving as multipath interferences during ripeness detection. We identify the ripeness of fruit as one of the four levels: *unripen*, *half ripen*, *ripen* and *over ripen*. Experimental results show that FruitSense is highly effective in detecting fruit ripeness. It achieves an accuracy at over 90% in identifying the ripeness level of fruit.

The main contributions of our work are summarized as follows:
- We show that the WiFi signals can be utilized to capture the physiological changes of fruit for ripeness detection. Our system is non-destructive and low-cost. It allows reusing existing WiFi devices at homes without the need for additional sensors.
- We leverage the frequency diversity to accurately capture the physiological changes of the fruit due to ripening, and utilize the larger bandwidth at 5GHz to combat multipath for extracting multipath-independent signal components for ripeness detection. We further utilize MODWT to extract features over multiple frequencies for identifying the level of fruit ripeness.
- We conduct extensive experiments in different multipath environments with two typical fruits (i.e., kiwi and avocado) under various conditions. Experimental results show that FruitSense achieves around 90% ripeness level detection accuracy.

## II. PRELIMINARIES

### A. Fruit Ripening

There exists two types of fruits according to the regulatory mechanisms underlying their ripening process: climacteric and non-climacteric fruits. Climacteric fruits, such as kiwi, avocado, apple and banana will continue to ripen after the fruit has left the plant; however, non-climacteric fruits, such as grape, orange and pineapple stop the ripening process the minute they leave the plant [36]. In this work, we focus on sensing the ripeness of climacteric fruits, as we do not have the access to the non-climacteric fruits that are on the plant. Nevertheless, our system can be extended to sense the maturity of non-climacteric fruits during the growth and development process while they are on the plant.

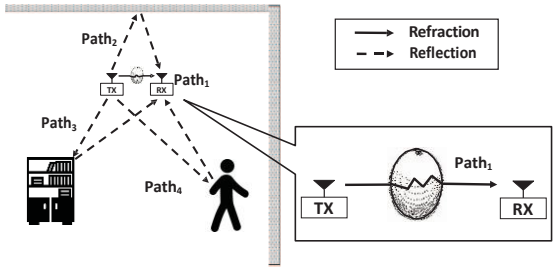 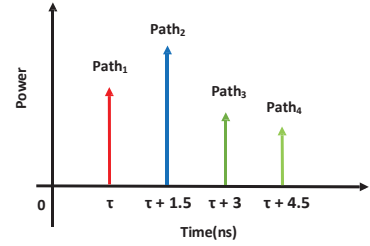

Fig. 1. Multipath propagations in an indoor environment during fruit ripeness sensing.

Fig. 2. Combining all the channels at 5GHz provides a power delay profile with sufficient resolution to differentiate multipath propagations.

Fruit ripening is a highly co-ordinated and an irreversible phenomenon involving a series of physiological and organoleptic changes [36]. For example, some fruits, such as banana and apple, come in a wide array of colors that change throughout their ripening process, with the brightest colors often occurring when the fruit is optimally ripened. Such color changes enable people to predict the ripeness level based on external visual inspection. By contrast, some other fruits, like kiwi and avocado, do not exhibit obvious organoleptic changes during the ripening process. It is hard for ordinary people to tell the ripeness level based on organoleptic testing. We thus experience with the ripeness detection of kiwi and avocado in this work. We seek to sense the fruit ripeness based on their physiological changes, instead of organoleptic changes.

Existing work uses the physiological changes such as oil context, dry matter and moisture content to determine the ripeness of avocado [31]. Specifically, oil content and dry matter increase during development and continue to change during ripening [31]. As oil content increases, the moisture content decreases by the same amount, so that the total percentage of oil and moisture content remains constant. Similarly, the content change within dry matter can be used to track the ripeness of kiwi fruit [36]. In particular, dry matter of kiwi is mainly comprised by the carbohydrates and starch, which gradually transform to soluble solids such as sugar content during the ripening. The percentage of carbohydrates and starch thus could be used as an indicator of fruit ripeness.

*B. WiFi Sensing based Approach*

Figure 1 illustrates a typical experimental setup of sensing fruit ripeness using a pair of WiFi devices in an indoor environment. The testing fruit is placed in between a pair of closely spaced transmitter and receiver, thus blocking the line-of-sight (LOS) signal propagation. We rely on the received signal component that directly travels through the fruit to sense its physiological change. The phenomenon where radio waves travel through and modulated by the fruit is commonly referred as refraction, which describes the signal passes from one medium to another. In our case, it is the WiFi signal travels from air to fruit, and then from fruit to air.

To quantify the effect of the refraction, we could leverage the concept of permittivity, which is a measure of how an electric field affects, and is affected by, a dielectric medium (i.e, the fruit in our case). In particular, the complex relative permittivity $\varepsilon^*$ of a material to that of free space in frequency domain can be described as following [27]:

$$\varepsilon^* = \varepsilon^I - j\varepsilon^{II} \quad (1)$$

The real part $\varepsilon^I$ is referred to as the dielectric constant, which describes the ability of the material to store energy when it is exposed to an electric field. The imaginary part $\varepsilon^{II}$ is referred to as dielectric loss factor, and $j = \sqrt{-1}$. The dielectric loss factor influences both energy attenuation and absorption, and is commonly used to describe the ability of the material to dissipate electrical energy as heat.

As fruit ripening involves a series of physiological changes, fruit at different ripeness levels results in different dielectric constants and loss factors. The permittivity of the fruit thus could be used as an indicator of its internal quality. Indeed, there exists prior work on using dielectric spectroscopy to measure the dielectric properties of fruit and vegetable for internal quality analysis[25].

As we use the off-the-shelf WiFi devices, we turn to exam how the received signal changes due to the changed physiological compounds of the fruit. Specifically, when the WiFi signal travels through the fruit, the electric field strength decreases with the distance from its surface. To quantify such an effect, the attenuation factor $\alpha$, which depends on the dielectric properties of the fruit, could be leveraged. It is given by:

$$\alpha = \frac{2\pi}{\lambda_0} \left( \frac{1}{2} \varepsilon_I \left( \sqrt{1 + \frac{\varepsilon^{II}}{\varepsilon^I}^2} - 1 \right) \right)^{1/2} \quad (2)$$

where $\lambda_0$ is the free-space wavelength of the WiFi signal [27]. Based on Equation 2, we know that the fruits at different ripeness levels lead to different attenuation factors, which could be measured by analyzing the received signals. Therefore, instead of using dedicated dielectric spectroscopy, we leverage the received signal that directly travels through the fruit for ripeness detection.

Excepting for the controllable experimental settings such as the location of the fruit and the distance between the transmitter and the receiver, the sizes of the fruit could be slightly different, which may affect the signal attenuation as well. Note that from Equation 2, we can observe that the WiFi signals at different frequencies (wavelengths) result in different attenuation factors as well. We thus can leverage the frequency diversity by probing the fruit at multiple channels, and then analyze the relative changes between multiple frequencies to mitigate the impact of fruit size.

*C. Challenges*

While the intuition is simple, there are significant challenges to accurately extracting the received signal that directly travels through the fruit. As shown in Figure 1, besides the signal refraction, there exists signal reflections. The signals that reflected from the walls, furniture, and human body will be combined with the signal travels through the fruit at the receiver. This leads to the fact that the measured signals at the receiver mainly reflect the multipath environments. To make the system robust to the multipath environment, we need to separate the signal component that travels through the fruit from the reflected ones.

Intuitively, this can be done by leveraging the power delay profile, which gives the power intensity of received signals as a function of propagation delay. By performing Inverse Fast Fourier Transform (IFFT) of the received signal measurements, we are able to extract the first arriving signal, which travels through the shortest path among all the paths from the transmitter to the receiver. However, the widely used bandwidth of a WiFi channel is either 20MHz or 40MHz, which results in a power delay profile with resolution at either 50ns or 25ns. Given that the wireless signal travels at the speed of light, such resolutions correspond to distance resolutions of 15m and 7.5m, respectively. As the size of a typical room is several meters by several meters, a majority of the reflected signals have path lengths smaller than 15m or 7.5m. Therefore, the obtained first arriving signal from the power delay profile based on each channel is still a mixture of the signals that travel through the fruit and multipaths. Therefore, simply performing IFFT on the signal measurements at each WiFi channel provides insufficient time or distance resolution for extracting the signal traveling through the fruit.

As WiFi spans multiple channels at both 2.4GHz and 5GHz, we propose to probe the fruit at all available channels of 5GHz with larger bandwidth. Combining all the channels at 5GHz (i.e., from 5.18GHz to 5.825GHz) brings over 600MHz bandwidth, which corresponds to a 1.5ns power delay profile resolution or to a 0.45 meters distance resolution. Such a resolution is sufficient for us to separate the signal traveling directly through the fruit from the reflected ones, as shown in Figure 2. Given the signal measurements at each channel, we then stitch these measurements together to derive a fine-grained power delay profile for multipath removal.

One problem with that stitching lies in the fact that the usable WiFi channels at 5GHz are unequally and non-contiguous spaced due to FCC regulation, as shown in Figure 4. For example, the channels from 120 to 128 are partially occupied by the weather radar usage in the US, and different countries apply their own regulations. Venders usually disable some of the 5GHz channels in compliance with the regulations of different countries before shipping the WiFi NICs. We thus cannot simply use IFFT, which only works for uniformly-spaced frequency measurements. To overcome this issue, we adopt the inverse non-uniform Discrete Fourier Transform (NDFT), which is capable of deriving a fine-grained power delay profile from non-uniformly spaced channels at 5GHz.

Another challenge lies in the use of off-the-shelf WiFi devices. Although the WiFi NIC employs sampling frequency offset (SFO) and carrier frequency offset (CFO) correctors to compensate the frequency offset errors, there still exists significant residual synchronization errors [20]. Figure 3 depicts the raw phases of two packets extracted at the NICs for the same multipath channel. Figure 3 (b) and (c) illustrate the derived power delay profiles with the raw phases presented in Figure 3 (a). Ideally, these two derived power delay profiles should be identical as they reflect the same multipath channel. However, Figure 3 (b) and (c) display the opposite observation: one profile exhibits LOS propagation while the other shows no LOS propagation. In particular, Figure 3 (b) shows the first arriving signal has the strongest power, whereas Figure 3 (c) depicts the first arriving signal is weaker than the reflected ones. To correct such an error, we utilize the fact that the power delay profiles derived from different WiFi channels should be identical as long as the probe packets were sent out within a coherence time. We thus could further compensate the frequency offset errors by searching for the best phase compensation that minimizes the differences in derived power delay profiles across multiple channels.

III. SYSTEM DESIGN

*A. System Overview*

Our system uses a pair of WiFi devices to sense the physiological compounds of fruit for ripeness detection. The major advantage of our method is that it is non-destructive and allows users to reuse existing WiFi devices without additional sensors. Figure 5 illustrates the flow of our system. It first probs the fruit with WiFi signals hopping through all usable channels at 5GHz. The system then collects the sampled channel frequency response in the form of Channel State Information (CSI) including the information of phase and amplitude. The CSI measurements are reported by the WiFi NIC at the receiver.

The CSI measurements are then preprocessed to calibrate both the phase and amplitude errors. In particular, the raw phase contains residual synchronization errors, which are composed of two types of errors: linear errors with respect to the subcarrier indexes and the constant errors across subcarriers. Our system calibrates the linear phase errors by searching for an optimum phase compensation value that minimizes the differences in the derived power delay profiles across multiple channels. The constant phase errors and amplitude errors could be mitigated by averaging the measurements of multiple packets that received in the coherence time window.

Given the calibrated CSI measurements, our system performs multipath removal for extracting the signal that directly travels through the fruit. Our system stitches the CSI measurements of all the channels together to enlarge the bandwidth for improving the resolution of the power delay profile. The inverse NDFT technique is used to overcome the problem of the unequal and non-contiguity spacing of 5GHz channels. Our system then identifies and extracts the signal directly travelling through the fruit for ripeness detection.

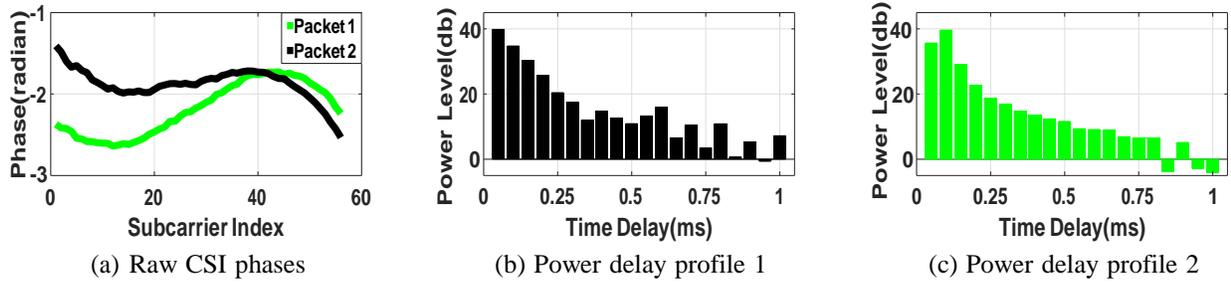

Fig. 3. Raw phases of two CSI measurements and the corresponding derived power delay profiles.

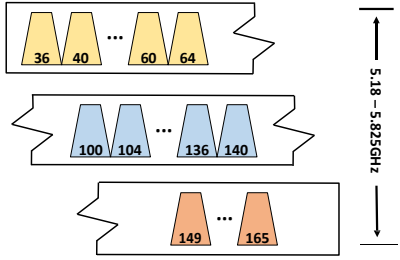

Fig. 4. The usable WiFi channels at 5GHz are non-uniformly spaced.

At last, our system identifies the fruit ripeness level by measuring the similarity between the extracted features and the pre-built fruit ripeness profiles. The MODWT is used to extract features over multiple channels and the cross correlation is used to compare the features against the ripeness level profiles in the library. The testing fruit is identified as one of the following ripeness levels: *unripen*, *half ripen*, *ripen* and *over ripen*.

### B. CFR Sampling

In an indoor environment, the signal undergoes multipath propagation. Assuming there are $L$ different paths, the signal attenuation and delay on $l$th path is $\alpha_l$ and $t_l$, respectively. The channel frequency response $h(f)$ can be described as following [15]:

$$h(f) = \sum_{l=0}^{L} \alpha_l e^{-j2\pi f t_l}, \quad (3)$$

where $f$ represents the center-frequency.

With 802.11n/ac systems, the WiFi NICs track fine-grained channel state information, which is a sampled version of the channel response including both phase and amplitude information. In particular, on the standard 20MHz WiFi channel, it measures the amplitude and phase for each of the 56 orthogonal frequency-division multiplexing (OFDM) subcarriers. With wider 40MHz channels, CSI measurements are available for 128 subcarriers. In our work, we utilize all available 20MHz WiFi channels at 5GHz with off-the-shelf WiFi NIC (i.e., Atheros AR9580). There are in total 21 channels usable in the Atheros NICs. To ensure that the channel hopping goes through all the channels within the coherence time, the hopping delay is set as 0.25ms. Note that the coherence time in typical indoor environment is at around 300ms [15]. This allows us to collect multiple packets at each channel within the coherence time. The following nominal matrix describes the collected CSI measurements during the ripeness sensing process:

$$\begin{bmatrix} \mathbf{csi}_{1,1} & \mathbf{csi}_{1,2} & \cdots & \mathbf{csi}_{1,q} & \cdots \\ \mathbf{csi}_{2,1} & \mathbf{csi}_{2,2} & \cdots & \mathbf{csi}_{2,q} & \cdots \\ \vdots & \vdots & & \vdots & \vdots \\ \mathbf{csi}_{p,1} & \mathbf{csi}_{p,2} & \cdots & \mathbf{csi}_{p,q} & \cdots \\ \vdots & \vdots & & \vdots & \vdots \end{bmatrix}, \quad (4)$$

where $\mathbf{csi}_{p,q}$ is the CSI complex vector sampled at the $p^{th}$ packet of the $q^{th}$ channel. Each CSI complex vector can be reported as a vector in frequency domain and represented as:

$$\mathbf{csi}_{p,q} = [csi_{p,q}^1, csi_{p,q}^2, \cdots, csi_{p,q}^K], \quad (5)$$

where $K$ is the total number of subcarriers reported at each channel per packet (i.e., K = 56 for 20MHz channel) and $csi_{p,q}^k$ is the sampled channel response at the $k$th subcarrier.

Since the CSI measurements are extracted by sampling the channel frequency response, the raw CSI measurements incur significant distortions due to the hardware limitations of off-the-shelf WiFi NICs. Such distortions or errors are mainly caused by clock unsynchronization. Thus, the reported CSI measurements with respect to the genuine channel response can be written as following:

$$\mathbf{csi}_{p,q}^k = |\mathbf{h}_{p,q}^k| e^{-jk\phi_l + \phi_c}, \quad (6)$$

where $\mathbf{h}_{p,q}^k$ is the frequency response caused by channel propagation, $\phi_l$ denotes the slope of the linear phase error and $\phi_c$ is the constant phase shift error. The reported CSI phase measurement $\angle \mathbf{csi}_{p,q}^k$ can be further represented as:

$$\angle \mathbf{csi}_{p,q}^k = \angle_{p,q}^k + k\phi_l + \phi_c, \quad (7)$$

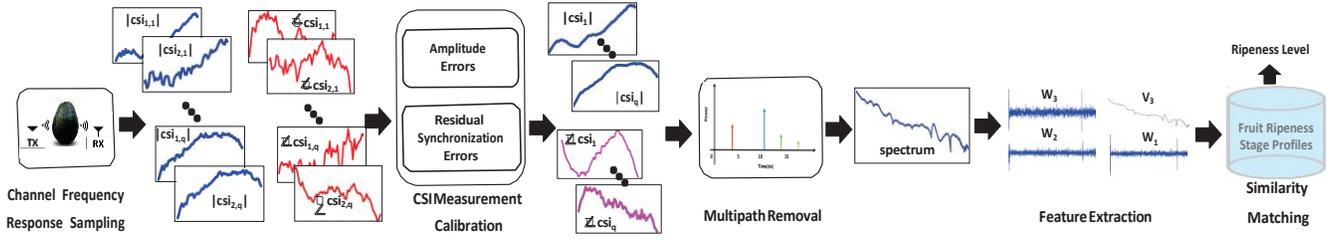

Fig. 5. System Overview.

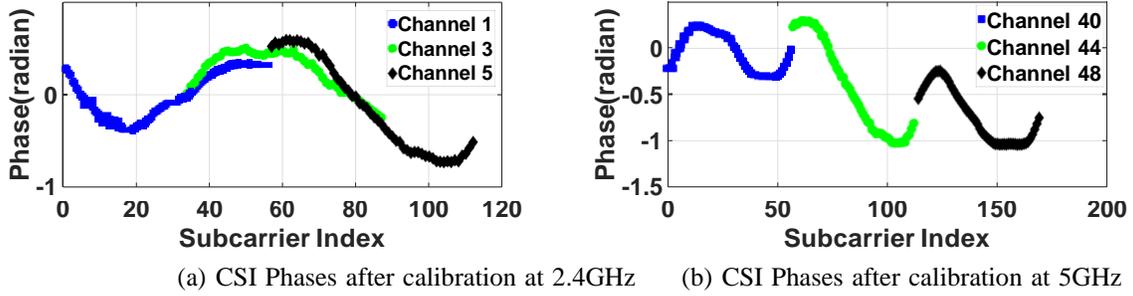

(a) CSI Phases after calibration at 2.4GHz   (b) CSI Phases after calibration at 5GHz

Fig. 6. Calibrated CSI phases for three overlapping channels at 2.4GHz and three channels at 5GHz.

where $\angle \mathbf{h}_{p,q}^k$ represents the phase rotation due to signal propagation. Meanwhile both $\phi_l$ and $\phi_c$ represent the residual synchronization errors of phase.

*C. Data Calibration*

As our system relies on the CSI measurements to derive fine-grained power delay profile for multipath removal, we first discover how the errors presented in the CSI measurements affect the derived power delay profile. First, different from the phase errors, the error of the amplitude doesn't affect the overall shape (or energy distribution with respect to time delay) of the power delay profile. The amplitude error is mainly caused by the power amplifier uncertainty. As the amplitude error is discovered to follow Gaussian distribution [23], it could be mitigated by averaging CSI measurements from multiple packets that within the coherence time. Among different types of phase errors, the constant phase error $\phi_c$, is caused by the residual central frequency offset. It does not affect the energy distribution with respect to time delay in the power delay profile as well. This is because the constant phase error is frequency independent. When performing IFFT, the phase shift $\phi_c$ would cause a constant rotation term for all the attenuation at different propagation delays. Thus, the derived power delay profile would stay the same after such a shift. Based on this observation, we select a reference channel from all the available channels and mitigate $\phi_c$ through correcting the phase difference from each channel according to the reference.

The power delay profile distortion presented in Figure 3 is mainly caused by the the linear phase error $\phi_l$, which is a frequency dependent error. We next detail how we correct the linear phase error $\phi_l$ by using the CSI measurements across multiple channels. The linear phase error $\phi_l$ is mainly caused by imperfectly synchronized clock at both transmitter and receiver ends. Even after phase compensation at WiFi NICs, there are still residual synchronization errors exist. In particular, it can be further divided into two components:

$$\phi_l = \phi_d + \phi_s, \quad (8)$$

The packet boundary detection could introduce uncertain time shift, which causes the phase error $\phi_d$. Such time shifts in reported CSI measurements are observed to follow the Gaussian distribution with zero mean [40]. Our experimental data also confirms such an observation. Furthermore, the error $\phi_d$ stays the same across subcarriers within a channel but changes across different channels. Because $\phi_d$ follows Gaussian distribution with zero mean, we are able to mitigate the error $\phi_d$ by averaging multiple CSI phase measurements collected at each channel.

The second component of linear error $\phi_s$ is introduced by the residual sampling frequency offset due to the unsynchronized clock at the transmitter and the receiver. Although the clock skews between different transmission pairs vary significantly, the skew remains constant for a few minutes for the same pair of transmitter and receiver [21]. Since our system probs the fruit with a single pair of WiFi device within one second, the linear error $\phi_s$ would remain constant across different packets and channels. To remove $\phi_s$, we utilize the fact that the power delay profiles derived from different WiFi channels should be identical, if the probe packets were sent out within the coherence time. This is because the measurements of these packets from multiple channels reflect the same multipath environment. The phase error $\phi_s$ calibration thus could be formulated as an optimization problem, in which we search for the best phase compensation value that minimizes

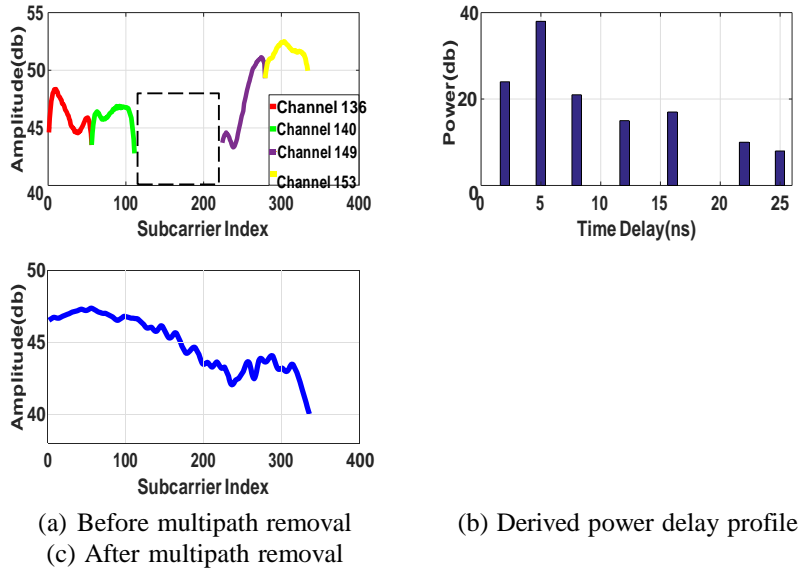

(a) Before multipath removal  (b) Derived power delay profile
(c) After multipath removal

Fig. 7. The CSI amplitude for four channels with gap at 5GHz before and after multipath removal.

the differences in derived power delay profiles across multiple channels.

Given the channel response, the power delay profile **g** at channel $q$ can be derived using IFFT:

$$\mathbf{g}_q = \sum_{l=1}^{L} a_l \delta(t - t_l), \quad (9)$$

where $l$ denotes the sequence number of total $L$ multipath channels, $a_l$ and $t_l$ are the amplitude and signal propagation time delay of $lth$ path, $\delta(t)$ is the Dirac delta function.

After removing phase errors $\phi_c$ and $\phi_d$ and assuming the the phase compensation for $\phi_s$ is $\phi_s^I$ Equation 7 can be rewritten as:

$$\angle \bar{\mathbf{csi}}_{p,q}^{k} = \angle_{p,q}^{k} + k\phi_s - k\phi_s^I. \quad (10)$$

By gradually changing the value of $\phi_s^I$, different power delay profiles can be derived for each channel. The optimum phase compensation $\phi_s$ can be found when the differences in derived power delay profiles across multiple channels are minimized. In particular, it can be formulated as an optimization problem as following:

$$\min_{\phi_s^I} \sum_{q,q^I=1}^{Q} \|\mathbf{g}_q(\phi_s^I) - \mathbf{g}_{q^I}(\phi_s^I)\|_2, \quad q \neq q^I, \quad (11)$$

where $\mathbf{g_q}(\phi_s^I)$ denotes the derived power delay profile at the $qth$ channel and $Q$ is the total number of channels. By obtaining an optimum value of $\phi_s^I$, we are able to remove the linear phase error $\phi_s$ from the CSI phase measurements.

Figure 6 shows the phases across different channels after calibrating $\phi_s$. As shown in Figure 6(a), after calibration, the phases of the overlapped subcarriers at three different 2.4GHz band channels now demonstrate similarity and consistency. Thus we can pick either one channel as reference to stitch them together. Figure 6(b) shows three non-overlapping channels at 5GHz band after phase calibration.

### D. Multipath Removal

Multipath removal leverages a fine-grained power delay profile to extract the signal that directly travels through the fruit for ripeness detection. Since a power delay profile derived from one single WiFi channel provides insufficient time delay resolution, our system stitches all the available channels at 5GHz together to improve the resolution for extracting the signal that travels through the shortest path.

Due to the regulations at many countries, the usable channels at 5GHz are unequally and non-contiguous spaced. For example, the usable channels at 5GHz on the Atheros NICs are separated into four segments: from 5.18GHz to 5.32 GHz, from 5.5GHz to 5.58 GHz, from 5.66GHz to 5.7 GHz, and from 5.745GHz to 5.825GHz. The venders disabled the rest channels in compliance with the local regulations. We thus cannot simply use IFFT for power delay profile derivation, as it only works for uniformly-spaced frequency measurements. Instead, we adopt inverse Non-uniform Discrete Fourier Transform (NDFT), which works for non-uniformly spaced channels. For fine-grained power profile derivation, given the calibrated CSI measurements from last step as:

$$\mathbf{CSI} = [\mathbf{CSI}_1, ..., \mathbf{CSI}_q, ...], \quad (12)$$

where $q$ denotes the $qth$ channel, we can formulate the inverse NDFT as following:

$$\min_{\mathbf{g}} \|\mathbf{CSI} - F\mathbf{g}\|^2, \quad (13)$$

where $\mathbf{g}$ represents the power delay profile that we search for, $F$ is Fourier matrix. The goal is to find an optimum solution of $\mathbf{g}$ that minimizes the difference between **CSI** and Fourier Transform of **g**.

Such an optimization problem can be viewed as an underdetermined system, which would yield several possible solutions. In order to pick out the best one, we need to add other constraints to filter out the less desired ones. To find such constraints, we look into the characteristic of signal propagation in indoor environment. Based on previous observations [10], even though multiple paths exist in typical indoor environment, only a few paths would dominate the signal propagation. It is because they travel through shorter paths and suffer less attenuations when comparing to longer propagations. By utilizing this observation, we add one constraint to the inverse NDFT: among all the solutions of **g** that satisfy Equation 13, our system favors the **g** with fewer dominating propagation paths.

To solve the Equation 13 with the constraint, we adopt the proximal gradient method that used to solve convex optimization problem [19]. To be more specific, our system takes **CSI** as input and computes the gradient of differentiable term in Equation 13. After obtaining several possible solutions, our system selects the one that with fewer dominating paths. Given the derived fine-grained power delay profile, we remove the components from multipath propagations and only keep the component that goes through the fruit directly. Then, we covert the trimmed power delay profile back to frequency domain for feature extraction.

Figure 7 shows the amplitude of the channels cover the same bandwidth before and after multipath removal process. As shown in Figure 7(a), four channels at 5GHz band cover 100MHz bandwidth with a 40MHz gap in between as highlighted by a black dot rectangle. After performing inverse NDFT, we are able to derive a fine-grained power delay profile shown in Figure 7(b), which indicated the lack of line of sight. Then we remove the multipath components from the power delay profile, only reserve the signal propagation affect by fruit and convert it back to frequency domain. The resulting spectrum is shown in Figure 7(c). We observe that the amplitude continue to decrease with increasing the frequency. It is consistent with Equation 2, where the attenuation factor increases with frequency. It means that a higher frequency would suffer a larger attenuation when propagating through the fruit.

### E. Feature Extraction and Ripeness Identification

Our system next uses Maximal Overlap Discrete Wavelet Transform (MODWT) to extract features based on the signals that directly travel through fruit at multiple frequencies (i.e., spectrum). Similar to discrete wavelet transform(DWT), MODWT analyzes the signals in both time and frequency domains by decomposing signals into successive approximation coefficients along with detailed coefficients. The approximation coefficients depict the large scale characteristic of change pattern, whereas the detailed coefficients capture small scale components that represent the fine details of the change pattern.

Unlike DWT, the MODWT is a nonorthogonal transform, which retains more information during decomposing process

| Fruit | Kiwi | | Avocado | |
|---|---|---|---|---|
| Ripeness | Vendor 1 | Vendor 2 | Vendor 1 | Vendor 2 |
| Unripen | 48 | 32 | 45 | 35 |
| Half Ripen | 43 | 31 | 43 | 33 |
| Ripen | 44 | 32 | 40 | 34 |
| Over Ripen | 42 | 30 | 44 | 34 |

Fig. 8. Volumes of fruits in our evaluation.

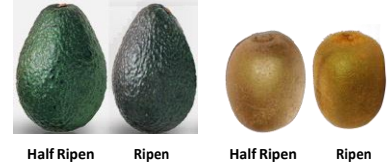

Half Ripen  Ripen   Half Ripen  Ripen

Fig. 9. Avocado and kiwi at two different ripeness levels.

that would be otherwise discarded by the DWT. Moreover, the MODWT is shift invariant, which means the decomposition output stay consistent across varies starting sample. This enables us to retrieve consistent features across a large number of channels. Given a signal **X(i)**, each level of MODWT coefficients are computed based on the following equations:

$$W_{j,i} = \sum_{l=0}^{L-1} h_{j,l} X_{i\_lmodN}, \\ V_{j,i} = \sum_{l=0}^{L-1} g_{j,l} X_{i\_lmodN}, \quad (14)$$

where $V_{j,i}$ is the approximation coefficient and $W_{j,i}$ is the detailed coefficient. Here $h_{j,l}$ and $g_{j,l}$ are the wavelet filter and scaling filter respectively. $j$ is the level of decomposition and $l = 1...L$ is the length of the filter. By applying MODWT to the extracted signal at each channel, we obtain both large scale and detailed features.

To identify the degree of ripeness, we utilize cross correlation to calculate the similarity between the extracted features and the ones in the profile library. Cross correlation measures the similarity based on the relative changes and is independent of translations and scaling in the amplitude. The ripeness level with the profile in the library that has the highest similarity with the testing fruit is then identified as the detected ripeness level.

## IV. PERFORMANCE EVALUATION

In this section, we describe our experimental setup and evaluate the performance of FruitSense in detecting the ripeness level of Kiwi fruit and Avocado under different multipath environments.

### A. Experiment Setup

We conduct experiments with two laptops (i.e., Dell LATITUDE E5540), each connecting with an external antenna.

The distance between two external antennas is about 20cm. The testing fruit is placed in the middle of the two antennas, thus blocking the line-of-sight propagation. Both laptops run Ubuntu 12.04 LTS and are equipped with the WiFi NICs of Atheros AR9580 for extracting CSI measurements [53]. The transmitter and receiver hop through all available 20MHz WiFi channels at 5GHz bands in an 802.11n network. There are total 21 available channels enabled by the Atheros AR9580 card. And they fall into four non-contiguous segments. The first segment is from 5.18GHz to 5.32 GHz (i.e., the channels from 36 to 64), whereas the second segment is from 5.5GHz to 5.58 GHz (i.e., the channels from 36 to 64). The third one and forth one are from 5.66GHz to 5.7 GHz (i.e., the channels from 132 to 140) and from 5.745GHz to 5.825GHz (i.e., the channels from 149 to 165), respectively.

Although the entire 5GHz has the bandwidth over 600MHz, we can only obtain 420MHz bandwidth from these four segments. We thus need to use inverse NDFT to derive the power delay profile for over 600MHz bandwidth based on unequally and non-contiguous spaced 21 channels. The channel hopping delay is set as 0.25ms. As the coherence time in typical indoor environment is about several hundreds milliseconds [15], we can collect packets across channels within coherence time as well as obtain multiple packets at each channel within coherence time. For each packet, we extract CSI for 56 subcarriers, which are equally distributed in a 20MHz channel.

We experiment with two types of commonly consumed fruits: kiwi fruit and avocado. The specific cultivars of avocado and kiwi fruit are Hass and Fuzzy respectively. Each type of fruit samples are purchased from two different vendors. Particularly, the production locations for kiwi fruits are USA and New Zealand for vender 1 and vender 2 respectively, whereas for avocado they are Mexico and USA for vender 1 and vender 2, respectively. The purchased fruits have four ripeness stages as commonly adopted by fruit industry: *unripen, half ripen, ripen* and *over ripen* [36], [52]. We use a spectrometer to log the ground truth of the fruit ripeness. For each ripeness stage and each type of fruit, we have around 75 fruits under testing. Figure 8 shows the detailed break down for each ripeness stage, vender, and type of fruit. Figure 9 gives one example of two ripeness stages of both kiwi and avocado. As we can see from Figure 9, both kiwi and avocado exhibit little external visual differences during ripening. To build the fruit ripeness profiles, we pre-select 10 fruits at each ripeness stage with various fruit sizes and average the features extracted from these fruits. All fruits are tested under the room temperature(23°$C$ to 26°$C$).

The experiments are conducted in two rooms at three locations, representing three different multipath environments. Figure 10 shows the layout of two rooms (i.e., one living room and one bedroom) and three locations. The bedroom has the size of 7 ft by 8 ft with one bed, one pair of table and chair. For the living room, it is 16 ft by 13 ft with regular living room furniture setup, such as dining table, book shelf, sofa, and TV. The bedroom environment represents a more compact space filled with furniture, while the living room setup describes a typical home environment with a larger space. To test the robustness of our system to the environment changes, we experience with people walking around during the ripeness detection. In particular, during the data collection, a person is walking around in the room to create interferences. The walking trajectory is shown in dash curve in Figure 10.

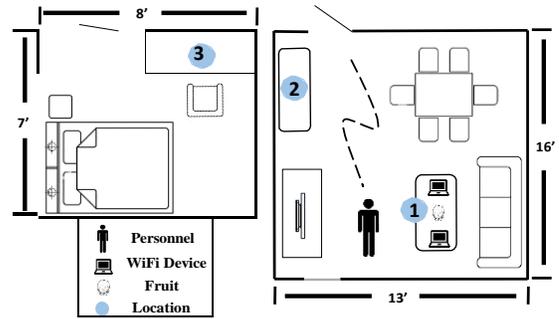

Fig. 10. Illustration of experiment setup.

We use both confusion matrix and detection accuracy to evaluate the performance of our system. For confusion matrix, each column represents the fruit ripeness level that was classified by our system and each row shows the ground truth of the fruit ripeness level. Each cell in the matrix corresponds to the fraction of ripeness level in the row that was classified as the ripeness level in the column. The detection accuracy is the percentage of the fruit that is correctly identified by our system.

*B. Overall Performance*

Figure 11 shows the confusion matrix of fruit ripeness detection for both kiwi and avocado. We observe that for both fruits, out system achieves overall detection accuracy over 90%. In particular, the overall detection accuracy for avocado is about 91%, whereas it is 90% for kiwi. Moreover, the half ripen and ripen fruit detection, compared to unripen and over ripen fruit detection, has a higher detection accuracy for both kiwi and avocado. Specifically, ripen fruit detection achieves 93% and 94% accuracy for kiwi and avocado respectively. This is due to the fact that the physiological changes at unripen and over ripen stages are at slower pace, whereas the changes at half ripen and ripen stages go through faster processes. Thus, more physiological changes of fruit at half ripen and ripen stages could be captured for ripeness detection. The above results show that our system could provide high accuracy in detecting fruit ripeness by using single pair of WiFi devices. The results also show that our system works with different multipath environments without location specific or environment specific calibration for ripeness profile. Still, the performance could be potentially improved by using additional pairs of WiFi devices or with multiple antennas.

We next study how the performance can be improved by increasing the sensed data from multiple days. Figure 12 depicts the overall detection accuracy when using the data

|          | Unripen | Half Ripen | Ripen | Over Ripen |
|----------|---------|------------|-------|------------|
| Unripen  |         | 0.06       | 0.03  | 0.03       |
| Half Ripen | 0.03  |            | 0.03  | 0.02       |
| Ripen    | 0.01    | 0.04       |       | 0.02       |
| Over Ripen | 0.02  | 0.05       | 0.06  |            |

(a) Kiwi

|          | Unripen | Half Ripen | Ripen | Over Ripen |
|----------|---------|------------|-------|------------|
| Unripen  |         | 0.06       | 0.03  | 0.03       |
| Half Ripen | 0.03  |            | 0.02  | 0.01       |
| Ripen    | 0.01    | 0.04       |       | 0.01       |
| Over Ripen | 0.02  | 0.04       | 0.05  |            |

(b) Avocado

Fig. 11. Confusion matrix of ripeness level detection for both kiwi and avocado fruits.

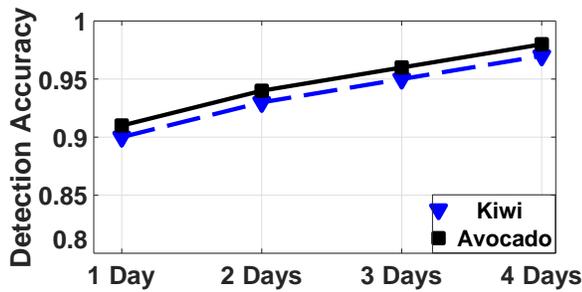

Fig. 12. Detection accuracy using the data collected from multiple days.

collected from a single day, two days, three and four days for both kiwi and avocado. For both kiwi and avocado, we observe that the performance increases when increasing the number of dates where the data are collected from. In particular, the detection accuracy increases from 93% to 97%, when the collected data are increased from two days to four days for kiwi. Similarly, the detection accuracy is increased from 94% to 98% for avocado when the data is increased from two days to four days. Overall, comparing to use data collected from one day, the system achieves over 8% improvement by combing the data from four days. Such an observation could benefit the venders or distributors who constantly tracking and monitoring the ripeness of fruit for much longer period than that of customers.

### C. Impact of Fruit Size

In last section, the ripeness profiles are built by averaging the features from the fruits with various sizes. We now evaluate

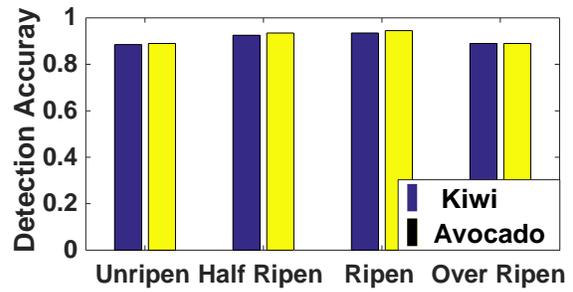

(a) Overall accuracy after separating fruits based on their sizes.

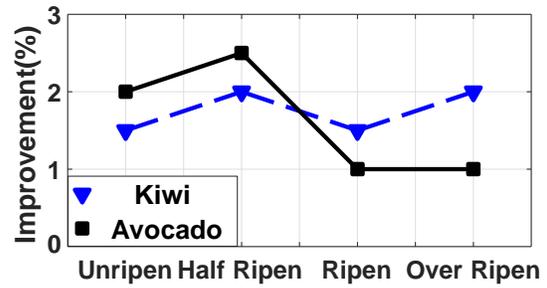

(b) Accuracy improvement with respect to mixed-size fruits.

Fig. 13. Detection accuracy and improvement after separating the small and the large size fruits.

the performance when the testing fruits have similar sizes as that of the fruits used to build ripeness profiles. In particular, the size of kiwi varies from 55mm to 63mm, whereas it is from 69mm to 85mm for avocado. We classify each type of fruit into two categories: small fruits and large fruits. Then, we sense the ripeness of small (large) fruits with the profiles built from the small (large) size fruits. Figure 13 shows the the overall accuracy for separating small and large size fruits and corresponding improvement with respect to mixed size fruits. We find that our system achieves slightly better performance at each ripeness level detection for both kiwi and avocado. In particular, it achieves an overall accuracy at around 92% for both kiwi and avocado. The overall improvement is at around 2% when comparing to that of the mixed fruit case. The results show that our system is not very sensitive to the fruit size, although a more fine-grained fruit profile could slightly improve the accuracy.

### D. Detailed Study on Multipath Environments

Although our overall performance in Section IV-B shows that our system works for different multipath environments, we further perform a more detailed study on whether the performance changes when using the ripeness profile built at one multipath environment to test the fruits at a different multipath environment. Specifically, we build the ripeness profiles when the fruits are at the location 1 in Figure 10. We then use such profiles to detect the fruits that placed at the location 2 (in the same room), and location 3 (in a different room). Figure 14(a) and Figure 14(b) present the performance of ripeness detection for kiwi and avocado, respectively. We observe that our system

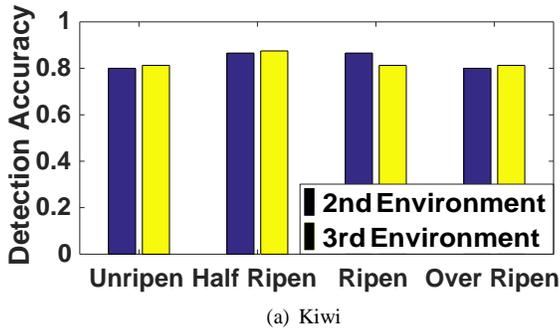

(a) Kiwi

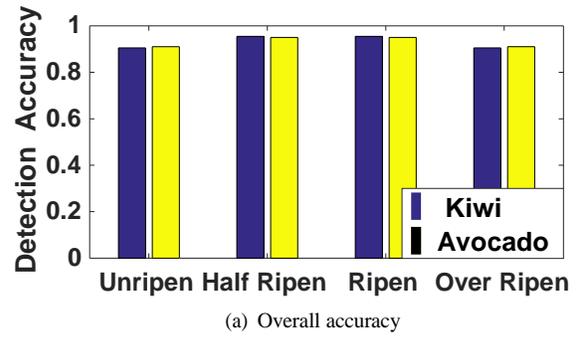

(a) Overall accuracy

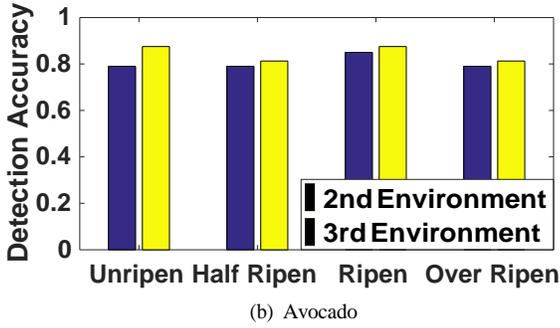

(b) Avocado

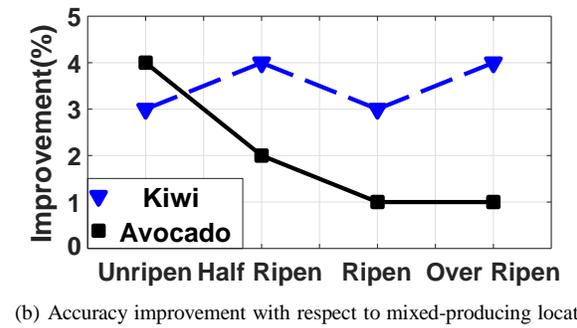

(b) Accuracy improvement with respect to mixed-producing location

Fig. 14. Detection accuracy under two multipath environments when using the ripeness profiles built under a different multipath environment.

Fig. 15. Detection accuracy and improvement after separating fruit based on producing location.

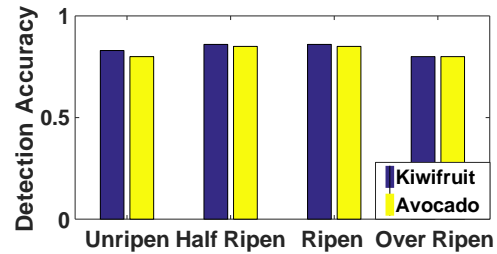

Fig. 16. Detection accuracy when people walking around during ripeness detection.

does not suffer from performance degradation due to the training and testing environments are under different multipath propagation. In particular, the overall detection accuracy still maintains at around 90% for both kiwi and avocado. These results further demonstrate the effectiveness of the multipath removal that leverages the larger bandwidth at 5GHz. It shows that our fruit ripeness detection system is location independent and environment independent. Once the profile is built, it can be applied to different environments without profile updating or calibration.

*E. Impact of Fruit Producing Location*

We next study the impact of production location of the fruit by separating each type of fruits based on their producing area. In particular, we build the ripeness profiles based on the producing area of each type of fruits. Then the ripeness profiles are used to test the fruits produced in the same area. The two producing areas for kiwi is USA and New Zealand, whereas they are USA and Mexico for avocado. The results are shown in Figure 15. We find that the detection accuracy has obvious improvement for both kiwi and avocado when compared to separating the fruits based on their sizes. Specifically, the overall accuracy is close to 94% for kiwi and at about 93% for avocado. The results show that the producing area of fruits play an important role when building the profile, even for the same type of fruit. This is because different producing areas result in different percentage of dry matter and moisture content, which is also discovered by existing work [33]. It suggests that a more fine-grained profile based on producing areas could improve the ripeness detection accuracy.

*F. Impact of People Movement*

Next we evaluate the detection accuracy of our system when there is people moving around during ripeness detection, as shown in Figure 10. Figure 16 depicts the detection accuracy at each ripeness level for both kiwi and avocado. We find our system still maintains good performance by comparing Figure 16 with Figure 11. It does not suffer obvious performance degradation due to the fact that the derived fine-grained power delay profile is able to filter out the signal reflections from the human body. This study demonstrates that our system is robust to the interferences from surroundings, such as people walking around.

## V. DISCUSSION

We conduct our experiments on kiwi fruit and avocado as they exhibit little visual differences during ripening. Besides kiwi and avocado, it is possible to extend our work by experimenting on larger size fruits such as: melon, pineapple and cantaloupe. Similar to kiwi and avocado, those types of fruits also show nearly no visual changes until they are over ripen. Such characteristic makes them the ideal testing subjects for our system. Moreover, we currently only focus on the post-harvest ripeness sensing, which mostly benefits customers and retainers. It is also possible to expand our work to pre-harvest ripeness tracking that provides wider range of assistance to farmers and fruit distributors.

Our current experiments only test on fruits purchased from two vendors with only two kinds of cultivars for each type of fruit. Each type of the fruits contains varies cultivars. Though sharing certain similarities, they all have their own distinctive physiological changes during ripening, which would yield different frequency responses when wireless signals passing through. It is thus highly desirable to build ripeness profile based on each type of cultivar. It is also worth mentioning our system only utilizes single pair of antennas. By adding additional antennas, the system should be able to capture more information about changes in ripeness level, and further improve its sensing accuracy.

## VI. RELATED WORK

There have been wide ranges of work utilizing specialized or commodity RF devices to perform human sensing. Those work can be further divided in three categories based on their applications: large scale movement tracking, small scale motion sensing and localization. For large scale movement tracking, E-Eyes and the other systems [48], [50], [14] are able to track daily activities such as cooking and bathing by using commodity WiFi devices, while WiSee [37] uses specialized RF device to sense whole home gestures including push and pull. Moreover, RF-Capture [5] utilizes multiple antennas to capture human figure through the wall. By using dedicated RF devices, Allsee [26] is capable of tracking hand movements.

For small scale motion sensing, WiKey [9] has the ability to sense typing on keyboard, whereas WiFinger[43] can recognize commonly used finger gestures. Moreover, WiHear utilizes multiple directional antenna to capture lip motion, and systems like Vital-Radio [30], [8], [14] are able to sense heartbeat and breathing. Besides motion tracking, many work have been proposed to enable localization utilizing RF signals [51]. Zhao and Patwari [55] proposed LSVRT to improve the localization accuracy by reducing the impact of variations caused by intrinsic motion, and Hillyard *et al.* develop new RSS-based crossing segment classifiers to localize outdoor boundary crossing. In addition, system such as Chronos [44] achieves decimeter-level localization with a single WiFi access point, and WiTrack and WiTrack 2.0 [6], [7] utilize RF body reflections to enable localization.

In general, the approaches for fruit ripeness detection can be divided into two categories: destructive and non-destructive methods. Abbott et al. [3] records the time-force curve as a small cube of sample is deformed between two plates at a constant speed. The data collected also known as texture profile, is used to estimate the degree of hardness in fruit. By measuring the sugar to acid ratio or total content of sugar or acid with refractometer, Jones and Scott [24] are able to track the ripeness level. Similar methods such as gas chromatograph [12] and mass spectrometer [39] can also be used to further analyze the chemical compounds of different fruits to identify different ripeness stages. Such methods however are destructive and compromise the intactness of sample and prevent it from future consumption.

For non-destructive methods, Peirs *et al.* [34] use NIR spectroscopy for post-harvest quality evaluation of apples. Peng and Lu [35] adopt multi-spectral and hyper-spectral imaging techniques to monitor apple samples by combining rich spectral information together. Imaging techniques such as X-ray computed tomography [22] and magnetic resonance imaging [11] have also been shown to be effective in fruit quality tracking. Moreover, ultrasonic signals have also been utilized to determine fruit maturity level [4] and detect the chilling injury in tomatoes [45]. The above approaches though provide non-destructive measuring, usually require expensive laboratory equipments, which are inaccessible to end-users. With recent development of optical technology, Das *et al.* [13] use portable spectrometer for non-destructive ripeness tracking by utilizing the UV fluorescence of Chlorophyll. Commercial devices such as SCiO [2] and Changhong H2 [1] can also achieve similar functionality using portable optical sensors. However, there are non-negligible costs incurred in purchasing dedicated spectral sensors.

## VII. CONCLUSION

This paper presents FruitSense, which is capable of sensing the ripeness of fruit by analyzing wireless signals traveling through the fruit. The proposed system, FruitSense, is non-destructive and only requires a pair of off-the-shelf WiFi devices. The insight is that the signals traveling through the fruit could capture the physiological changes associated with fruit ripening. Experimental results under different multipath environments show that FruitSense can detect the ripeness levels of both kiwi fruit and avocado with an accuracy over 90%. FruitSense builds on the growing interest within the wireless and mobile computing communities in using wireless signals for human sensing. The proposed work further expands the scope of human sensing to sensing the bio-information of fruit crops. We believe the implication of the work could be extended beyond classifying the fruit ripeness, particularly in quantifying the physiological compounds of fruit with more advanced signal process and machine learning techniques.

## REFERENCES

[1] Changhong. https://liliputing.com/2017/01/changhong-h2-smartphone-built-spectrometer.html, note = Accessed: 2017-03-29.
[2] Scio. https://www.consumerphysics.com/. Accessed: 2017-03-29.
[3] J. Abbott, D. Massie, and A. W. ADA. The use of a computer with an instron for textural measurements3. *Journal of Texture Studies*, 13(4):413–422, 1982.
[4] S. Adhimantoro and F. L. Gaol. Application of ultrasonic and fuzzy logic to determine fruit maturity level. *International Journal of Control and Automation*, 7(1):27–38, 2014.
[5] F. Adib, C.-Y. Hsu, H. Mao, D. Katabi, and F. Durand. Capturing the human figure through a wall. *ACM Transactions on Graphics (TOG)*, 34(6):219, 2015.

[6] F. Adib, Z. Kabelac, and D. Katabi. Multi-person localization via rf body reflections. In *NSDI*, pages 279–292, 2015.

[7] F. Adib, Z. Kabelac, D. Katabi, and R. C. Miller. 3d tracking via body radio reflections. In *NSDI*, volume 14, pages 317–329, 2014.

[8] F. Adib, H. Mao, Z. Kabelac, D. Katabi, and R. C. Miller. Smart homes that monitor breathing and heart rate. In *ACM CHI*, 2015.

[9] K. Ali, A. X. Liu, W. Wang, and M. Shahzad. Keystroke recognition using wifi signals. In *ACM MobiCom*, 2015.

[10] W. U. Bajwa, J. Haupt, A. M. Sayeed, and R. Nowak. Compressed channel sensing: A new approach to estimating sparse multipath channels. *Proceedings of the IEEE*, 98(6):1058–1076, 2010.

[11] P. Barreiro, C. Ortiz, M. Ruiz-Altisent, J. Ruiz-Cabello, M. E. Fernańdez-Valle, I. Recasens, and M. Asensio. Mealiness assessment in apples and peaches using mri techniques. *Magnetic Resonance Imaging*, 18(9):1175–1181, 2000.

[12] R. G. Buttery, R. Teranishi, R. A. Flath, and L. C. Ling. Fresh tomato volatiles. ACS Publications, 1989.

[13] A. J. Das, A. Wahi, I. Kothari, and R. Raskar. Ultra-portable, wireless smartphone spectrometer for rapid, non-destructive testing of fruit ripeness. *Scientific Reports*, 6, 2016.

[14] B. Fang, N. D. Lane, M. Zhang, A. Boran, and F. Kawsar. Bodyscan: Enabling radio-based sensing on wearable devices for contactless activity and vital sign monitoring. In *Proceedings of the 14th Annual International Conference on Mobile Systems, Applications, and Services*, pages 97–110. ACM, 2016.

[15] A. Goldsmith. *Wireless communications*. Cambridge university press, 2005.

[16] F. Harker, J. Maindonald, and P. Jackson. Penetrometer measurement of apple and kiwifruit firmness: operator and instrument differences. *Journal of the American Society for Horticultural Science*, 1996.

[17] R. Harrill. Using a refractometer to test the quality of fruits and vegetables. *P. PUBLISHING, Ed.) Consulté le July*, 20:2010, 1998.

[18] P. Hillyard, A. Luong, and N. Patwari. Highly reliable signal strength-based boundary crossing localization in outdoor time-varying environments. In *Proceedings of the 15th International Conference on Information Processing in Sensor Networks*, page 6. IEEE Press, 2016.

[19] K. Hou, Z. Zhou, A. M.-C. So, and Z.-Q. Luo. On the linear convergence of the proximal gradient method for trace norm regularization. In *Advances in Neural Information Processing Systems*, pages 710–718, 2013.

[20] X. Hu, Y. Huang, and Z. Hong. Residual synchronization error elimination in ofdm baseband receivers. *ETRI journal*, 2007.

[21] S. Jana and S. K. Kasera. On fast and accurate detection of unauthorized wireless access points using clock skews. *IEEE Transactions on Mobile Computing*, 9(3):449–462, 2010.

[22] J.-A. Jiang, H.-Y. Chang, K.-H. Wu, C.-S. Ouyang, M.-M. Yang, E.-C. Yang, T.-W. Chen, and T.-T. Lin. An adaptive image segmentation algorithm for x-ray quarantine inspection of selected fruits. *Computers and electronics in agriculture*, 60(2):190–200, 2008.

[23] V. P. G. Jimenez, M.-G. Garcia, F. G. Serrano, and A. G. Armada. Design and implementation of synchronization and agc for ofdm-based wlan receivers. *IEEE Transactions on Consumer Electronics*, 50(4):1016–1025, 2004.

[24] R. Jones and S. Scott. Improvement of tomato flavor by genetically increasing sugar and acid contents. *Euphytica*, 32(3):845–855, 1983.

[25] J. Juansah, I. Budiastra, K. Dahlan, and B. Seminar. Electrical behavior of garut citrus fruits during ripening changes in resistance and capacitance models of internal fruits. *Int. J. Eng. Technol. IJET-IJENS*, 12(4):1–8, 2012.

[26] B. Kellogg, V. Talla, and S. Gollakota. Bringing gesture recognition to all devices. In *NSDI*, volume 14, pages 303–316, 2014.

[27] V. Komarov, S. Wang, and J. Tang. Permittivity and measurements. *Encyclopedia of RF and microwave engineering*, 2005.

[28] H. Liu, Y. Gan, J. Yang, S. Sidhom, Y. Wang, Y. Chen, and F. Ye. Push the limit of wifi based localization for smartphones. In *Proceedings of the 18th annual international conference on Mobile computing and networking*, pages 305–316, 2012.

[29] J. Liu, Y. Chen, Y. Wang, X. Chen, J. Cheng, and J. Yang. Monitoring vital signs and postures during sleep using wifi signals. *IEEE Internet of Things Journal*, 5(3):2071–2084, 2018.

[30] J. Liu, Y. Wang, Y. Chen, J. Yang, and X. Chen. Tracking vital signs during sleep leveraging off-the-shelf wifi. In *ACM MobiHoc*, 2015.

[31] L. S. Magwaza and S. Z. Tesfay. A review of destructive and non-destructive methods for determining avocado fruit maturity. *Food and bioprocess technology*, 8(10):1995–2011, 2015.

[32] B. M. Nicolaï, T. Defraeye, B. De Ketelaere, E. Herremans, M. L. Hertog, W. Saeys, A. Torricelli, T. Vandendriessche, and P. Verboven. Nondestructive measurement of fruit and vegetable quality. *Annual review of food science and technology*, 5:285–312, 2014.

[33] F. Ozdemir and A. Topuz. Changes in dry matter, oil content and fatty acids composition of avocado during harvesting time and post-harvesting ripening period. *Food Chemistry*, 86(1):79–83, 2004.

[34] A. Peirs, N. Scheerlinck, K. Touchant, and B. M. Nicolaı. Ph—postharvest technology: Comparison of fourier transform and dispersive near-infrared reflectance spectroscopy for apple quality measurements. *Biosystems Engineering*, 81(3):305–311, 2002.

[35] Y. Peng and R. Lu. Analysis of spatially resolved hyperspectral scattering images for assessing apple fruit firmness and soluble solids content. *Postharvest Biology and Technology*, 48(1):52–62, 2008.

[36] V. Prasanna, T. Prabha, and R. Tharanathan. Fruit ripening phenomena–an overview. *Critical reviews in food science and nutrition*, 2007.

[37] Q. Pu, S. Gupta, S. Gollakota, and S. Patel. Whole-home gesture recognition using wireless signals. In *ACM MobiCom*, 2013.

[38] S. Saevels, J. Lammertyn, A. Z. Berna, E. A. Veraverbeke, C. Di Natale, and B. M. Nicolaı. An electronic nose and a mass spectrometry-based electronic nose for assessing apple quality during shelf life. *Postharvest Biology and Technology*, 31(1):9–19, 2004.

[39] P. Schreier, F. Drawert, and A. Junker. Identification of volatile constituents from grapes. *Journal of Agricultural and Food Chemistry*, 24(2):331–336, 1976.

[40] M. Speth, S. A. Fechtel, G. Fock, and H. Meyr. Optimum receiver design for wireless broad-band systems using ofdm. i. *IEEE Transactions on communications*, 47(11):1668–1677, 1999.

[41] A. Sugiura, I. Kataoka, and T. Tomana. Use of refractometer to determine soluble solids of astringent fruits of japanese persimmon (diospyros kaki l.). *Journal of horticultural science*, 58(2):241–246, 1983.

[42] S. Tan and J. Yang. Fine-grained gesture recognition using wifi. In *2016 IEEE Conference on Computer Communications Workshops (INFOCOM WKSHPS)*, pages 257–258. IEEE, 2016.

[43] S. Tan and J. Yang. Wifinger: leveraging commodity wifi for fine-grained finger gesture recognition. In *ACM MobiHoc*, 2016.

[44] D. Vasisht, S. Kumar, and D. Katabi. Decimeter-level localization with a single wifi access point. In *USENIX NSDI*, 2016.

[45] B. E. Verlinden, V. De Smedt, and B. M. Nicolaı. Evaluation of ultrasonic wave propagation to measure chilling injury in tomatoes. *Postharvest biology and technology*, 32(1):109–113, 2004.

[46] C. Wang, X. Zheng, Y. Chen, and J. Yang. Locating rogue access point using fine-grained channel information. *IEEE Transactions on Mobile Computing*, 16(9):2560–2573, 2016.

[47] G. Wang, Y. Zou, Z. Zhou, K. Wu, and L. M. Ni. We can hear you with wi-fi! 2014.

[48] Y. Wang, J. Liu, Y. Chen, M. Gruteser, J. Yang, and H. Liu. E-eyes: device-free location-oriented activity identification using fine-grained wifi signatures. In *ACM MobiCom*, 2014.

[49] Y. Wang, J. Yang, H. Liu, Y. Chen, M. Gruteser, and R. P. Martin. Measuring human queues using wifi signals. In *Proceedings of the 19th annual international conference on Mobile computing & networking*, pages 235–238, 2013.

[50] B. Wei, W. Hu, M. Yang, and C. T. Chou. Radio-based device-free activity recognition with radio frequency interference. In *Proceedings of the 14th International Conference on Information Processing in Sensor Networks*, pages 154–165. ACM, 2015.

[51] B. Wei, A. Varshney, N. Patwari, W. Hu, T. Voigt, and C. T. Chou. drti: Directional radio tomographic imaging. In *Proceedings of the 14th International Conference on Information Processing in Sensor Networks*, pages 166–177. ACM, 2015.

[52] R. Wills and J. Golding. *Postharvest: an introduction to the physiology and handling of fruit and vegetables*. UNSW press, 2016.

[53] Y. Xie, Z. Li, and M. Li. Precise power delay profiling with commodity wifi. In *ACM MobiCom*, 2015.

[54] J. Yang, Y. Ge, H. Xiong, Y. Chen, and H. Liu. Performing joint learning for passive intrusion detection in pervasive wireless environments. In *2010 Proceedings IEEE INFOCOM*, pages 1–9. IEEE, 2010.

[55] Y. Zhao and N. Patwari. Robust estimators for variance-based device-free localization and tracking. *IEEE Transactions on Mobile Computing*, 14(10):2116–2129, 2015.

[56] X. Zheng, H. Liu, J. Yang, Y. Chen, R. P. Martin, and X. Li. A study of localization accuracy using multiple frequencies and powers. *IEEE Transactions on Parallel and Distributed Systems*, 25(8):1955–1965, 2013.

[57] X. Zheng, J. Yang, Y. Chen, and H. Xiong. An adaptive framework coping with dynamic target speed for device-free passive localization. *IEEE Transactions on Mobile Computing*, 14(6):1138–1150, 2014.

[58] M. Zude-Sasse, I. Truppel, and B. Herold. An approach to non-destructive apple fruit chlorophyll determination. *Postharvest Biology and Technology*, 25(2):123–133, 2002.